\documentclass[twocolumn,tighten]{aastex7}
\pdfoutput=1 
\usepackage{amsmath,amstext}
\usepackage[T1]{fontenc}
\usepackage{newtxtext,newtxmath}
\usepackage[utf8]{inputenc}
\usepackage[figure,figure*]{hypcap}
\usepackage{enumitem}
\usepackage{bm}
\usepackage{graphicx}
\usepackage{lineno}
\usepackage{tikz}
\usepackage{mathtools}
\makeatletter
\@ifundefined{mleft}{%
  \let\ml\left
  \let\mr\right
}{%
  \let\ml\mleft
  \let\mr\mright
}
\makeatother

\newcommand{\Vone}{%
\mathord{\vcenter{\hbox{%
\begin{tikzpicture}[baseline=-0.6ex, scale=0.125]
  \draw[line width=0.8pt] (0,0) -- (0,-2) -- (1,0);
\end{tikzpicture}%
}}}}

\linenumbers
\input{hyperlink-year-only-natbib-patch}

\newcommand{\Mpc}{{{\rm Mpc}}}
\newcommand{\znr}{{z_{\rm nr}}}

\newcommand{\Beq}{\begin{equation}\begin{aligned}}
\newcommand{\Eeq}{\end{aligned}\end{equation}}
\newcommand{\dl}{\mathrm{d}}

\graphicspath{{./}{}}

\shorttitle{Warm, not Fuzzy}
\shortauthors{Nadler et al.}

\begin{document}

\title{Warm, not Fuzzy: Generalized Ultralight Dark Matter Limits from Milky Way Satellites}

\author[0000-0002-1182-3825]{Ethan~O.~Nadler}
\email{enadler@ucsd.edu}
\affiliation{Department of Astronomy \& Astrophysics, University of California, San Diego, La Jolla, CA 92093, USA}
\author[0000-0002-8742-197X]{Mustafa~A.~Amin}
\email{mustafa.a.amin@rice.edu}
\affiliation{Department of Physics and Astronomy, Rice University, Houston, TX 77005, USA}
\author[0000-0003-2229-011X]{Risa~H.~Wechsler}
\email{rwechsler@stanford.edu}
\affiliation{Kavli Institute for Particle Astrophysics and Cosmology and Department of Physics, Stanford University, Stanford, CA 94305, USA}
\affiliation{SLAC National Accelerator Laboratory, Menlo Park, CA 94025, USA}
\author[0000-0003-3808-5321]{M.~Sten Delos}
\email{mdelos@carnegiescience.edu}
\affiliation{Carnegie Observatories, 813 Santa Barbara Street, Pasadena, CA 91101, USA}
\author[0000-0001-5501-6008]{Andrew Benson}
\email{abenson@carnegiescience.edu}
\affiliation{Carnegie Observatories, 813 Santa Barbara Street, Pasadena, CA 91101, USA}
\author[0000-0002-3589-8637]{Vera Gluscevic}
\email{verag@ias.edu}
\affiliation{Department of Physics $\&$ Astronomy, University of Southern California, Los Angeles, CA 90007, USA}
\affiliation{Institute for Advanced Study, 1 Einstein Drive, Princeton, NJ 08540, USA}
\affiliation{Center for Computational Astrophysics, Flatiron Institute, 162 5th Avenue, New York, NY, 10010, USA}

\correspondingauthor{Ethan~O.~Nadler}
\email{enadler@ucsd.edu}

\begin{abstract}

We generalize lower limits on the dark matter (DM) particle mass $m$ derived from Milky Way (MW) satellite galaxy abundances to scenarios in which DM is an ultralight scalar field produced with a field power spectrum peaked at a subhorizon wavenumber $k_*$. In these models, the DM field free-streams similarly to warm DM while also exhibiting significant small-scale wave interference effects. The resulting dimensionless density power spectrum shows two effects: (i) free-streaming suppression at $k_{\rm fs}\sim k_{\rm eq}/[(k_*/a_{\rm eq}m)\ln(a_{\rm eq}m/k_*)]$; (ii) Poisson-like enhancement related to wave interference at $k\gtrsim10^{-2}k_*$, which saturates near the Jeans scale $k_{\rm J}\sim k_{\rm eq}/(k_*/a_{\rm eq}m)$. Comparing these predictions with established constraints on a free-streaming cutoff in the linear matter power spectrum from the MW satellite population and assuming that warm ultralight DM does not change the form of the galaxy--halo connection relative to cold DM, we obtain $m>6\times10^{-18}\,{\rm eV}\,(k_*/10^4\,{\rm Mpc}^{-1})$ for $k_*>10^4\,{\rm Mpc}^{-1}$ at 95\% confidence. For smaller $k_*$, Poisson-noise enhancement on MW satellite scales weakens the constraint, yielding $m>6\times10^{-18}\,{\rm eV}\,(k_*/10^4\,{\rm Mpc}^{-1})^2$ for $k_*<10^4\,{\rm Mpc}^{-1}$ at 95\% confidence.

\end{abstract}

\keywords{\href{http://astrothesaurus.org/uat/353}{Dark matter (353)}; 
\href{http://astrothesaurus.org/uat/574}{Galaxy abundances (574)};
\href{http://astrothesaurus.org/uat/1787}{Warm dark matter (1787)}
}

\section{Introduction}
\label{sec:intro}

When and how was dark matter (DM) produced? The answer to this question remains unknown and is central to understanding the microphysical nature of DM. 

DM fields produced through processes with a finite subhorizon correlation length have an isocurvature Poisson-type enhancement in their density fluctuations on scales larger than the correlation length. Consistent with observations, the usual adiabatic density perturbations dominate on even larger scales. In these scenarios, the DM field lacks a significant zero-momentum mode and has large gradients, resulting in free-streaming that suppresses adiabatic density perturbations and inhibits the growth of the isocurvature enhancement below the classical Jeans scale~\citep{Amin221109775,Amin250612131,Amin250320881,Liu250401937,Liu240612970}. We emphasize that initial field configurations in this ``warm'' wave DM case are highly inhomogeneous, in contrast to cold fuzzy DM (FDM), where the DM field is homogeneous with small variations. 

These scenarios are of broad theoretical and observational interest. For example, most postinflationary DM production mechanisms yield the free-streaming suppression and Poisson enhancement described above if DM is an ultralight scalar (e.g., an open-string axion; \citealt{Allahverdi:2014ppa,Petrossian-Byrne:2025mto}). Thus, while cold FDM suppresses structure due to its astrophysically large Jeans length~\citep{Hu008506,Hui179504}, free-streaming of warm wave DM can suppress power on still larger scales, or smaller wavenumbers $k$ \citep{Amin221109775}. The classical Jeans scale also appears on larger length scales than the ``quantum'' Jeans scale and leads to a scale-dependent growth of the isocurvature Poisson-type enhancement \citep{Amin250612131}. 

Such ultralight DM can effectively free-stream like warm dark matter (WDM) and also exhibit significant density fluctuations related to wave interference; for recent related work, see \cite{Ling240805591}, \cite{Amin:2025ayf,Amin250612131,Amin250320881}, \cite{Amin:2025nxm}, \cite{Long241214322}, \cite{Liu250401937,Liu240612970}, \cite{Harigaya:2025pox}, \cite{Gorghetto:2025uls}, and \cite{Chathirathas:2025aan}. This free-streaming suppression does not require DM to be produced thermally. Indeed, a number of inflationary production mechanisms yield a field spectrum devoid of a significant zero-momentum mode \citep{Kolb231209042}, leading to similar phenomenology.\footnote{One relevant exception is light, minimally coupled scalar DM produced during inflation, where the density is dominated by the zero-momentum (homogeneous) field mode. This includes axion DM produced via the global misalignment mechanism (e.g., see \citealt{Marsh151007633} for a review).}

As a result of this free-streaming, the linear matter power spectrum, $P(k)$, can be suppressed in ultralight DM models in a manner similar to thermal-relic WDM~\citep{Bond161460,Bode0010389,Schneider11120330}. Small-scale structure measurements have placed stringent constraints on such suppression. For example, an analysis of the Milky Way (MW) satellite galaxy population observed by the Dark Energy Survey (DES) and Pan-STARRS1 (PS1), using data compiled by \cite{DrlicaWagner191203302}, found that $P(k)$ is not suppressed by more than $75\%$ relative to cold, collisionless dark matter (CDM) at the half-mode wavenumber $k_{\mathrm{hm}}=50~\mathrm{Mpc}^{-1}$ at $95\%$ confidence \citep[see also \citealt{Newton201108865,Dekker211113137}]{Nadler200800022}. This constraint assumed a thermal-relic WDM power spectrum but has recently been extended to more general $P(k)$ shapes~\citep{Nadler241003635,An241103431}.

Here, we combine this progress in constraining models that suppress $P(k)$ using MW satellite galaxies with the theoretical framework of \cite{Amin250612131, Amin250320881} to derive a lower limit on the mass $m$ of DM particles as a function of the comoving momentum $k_*$. This comoving momentum is the location of the peak in the DM field power spectrum before matter--radiation equality, and it determines the dominant field modes that contribute to the DM density and characteristic velocity. Larger values of $k_*$ imply a larger free-streaming length at fixed DM mass, with $k_{\rm fs}^{-1}\propto k_*/m$ to leading order.

Specifically, we calculate $P(k)$ for warm ultralight DM fields (see Figure~\ref{fig:T_DM}) and map the result to a recent $P(k)$ constraint from MW satellite abundances derived using the COZMIC simulations~\citep{An241103431,Nadler241003635,Nadler241213065}. In this way, we obtain lower limits on the DM mass as a function of $k_*$, ruling out $m<6\times 10^{-18}~\mathrm{eV}$ at $k_*>10^4~\mathrm{Mpc}^{-1}$ at $95\%$ confidence (see Figure~\ref{fig:constraint}). These limits are conservative because we require that $P(k)$ be strictly more suppressed than the ruled-out COZMIC model over the entire wavenumber range relevant for MW satellites. We also discuss how other small-scale structure probes, including the Ly$\alpha$ forest (e.g., \citealt{Amin221109775,Long241214322}), complement our limits on this class of DM models.

This Letter is organized as follows. In Section~\ref{sec:methods}, we describe our method for calculating $P(k)$ for ultralight DM in the presence of free-streaming effects, and how we map these predictions to $P(k)$ constraints from MW satellite galaxies. In Section~\ref{sec:results}, we derive our bounds on the particle mass $m$ as a function of $k_*$; we discuss our results in Section~\ref{sec:discussion} and conclude in Section~\ref{sec:conclusion}.

We use cosmological parameters of~$h = 0.7$, $\Omega_{\rm m} = 0.286$, $\Omega_{\rm b} = 0.047$, $\Omega_{\Lambda} = 0.714$, and $\sigma_8=0.82$ for consistency with the \cite{An241103431} COZMIC II simulations, and we define halo masses using the \cite{Bryan9710107} virial overdensity, which corresponds to $\Delta_{\rm vir}\simeq 99.2$ at $z=0$ in our fiducial cosmology. Throughout, we set $\hbar=c=1$.

\section{Linear Matter Power Spectrum}
\label{sec:methods}

\subsection{Model}
To calculate the linear matter power spectrum $P(z,k)$ in a scenario where DM density perturbations are sourced by an underlying field spectrum peaked at $k_*$, we define the transfer function
\begin{align}
    T_{\mathrm{dm}}^2(y,k;m,k_*) 
    &\equiv \frac{P_{\mathrm{dm}}(y,k;m,k_*)}{P_{\mathrm{cdm}}(y,k)},\nonumber\\
    &=\frac{P^{(\mathrm{ad})}_{\mathrm{dm}}(y,k;m,k_*)+P^{(\mathrm{iso})}_{\mathrm{dm}}(y,k;m,k_*)}{P_{\mathrm{cdm}}(y,k)},
\end{align}
where $y=a/a_{\rm eq}=(1+z_{\rm eq})/(1+z)$, $P_{\mathrm{dm}}(y,k;m,k_*)$ is the linear matter power spectrum in our DM model, and $P_{\mathrm{cdm}}(y,k)$ is the CDM power spectrum. 

In terms of initial conditions at some $y_0\ll 1$, we write
\begin{align}\label{eq:power}
P_{\rm cdm}(y,k)&=P_{\rm cdm}(y_0,k)\mathcal{T}^2_{\rm cdm}(y,y_0,k)\nonumber\\
P^{({\rm ad})}_{\rm dm}(y,k)&=P_{\rm cdm}(y_0,k)\mathcal{T}^{2(\rm ad)}_{\rm dm}(y,y_0,k;m/k_*)\nonumber\\
P^{({\rm iso})}_{\rm dm}(y,k)&=P^{(\rm iso)}_{\rm dm}(y_0,k)\left[\mathcal{T}^{2(\rm iso)}_{\rm dm}(y,y_0,k;m,k_*)-1\right]\nonumber\\
\end{align}
where $\mathcal{T}^{(\rm iso, ad)}_{\rm dm}$, which evolve the initial density power spectrum, are obtained using the framework described in \cite{Amin250612131}.\footnote{The code we use to evaluate these power spectra is publicly available at \url{https://github.com/delos/warm-structure-growth} \citep{delos_2025_17064722}. In Equation~(\ref{eq:power}), we subtract the intrinsic Poisson noise term, $P^{(\rm iso)}_{\rm dm}(y_0,k)$, since it is not associated with gravitational clustering and hence would not contribute to halo formation.} We summarize the key formulae and results related to this evolution in Appendix \ref{App:PS}. The main input for calculating the evolution is the field power spectra at $y=y_0$. For concreteness, we take this to be shaped like a Maxwell--Boltzmann distribution, i.e., a Gaussian with width determined by $k_*$.\footnote{We have tested a variety of additional velocity distribution shapes including uniform, exponential, and parabolic. This choice negligibly impacts our results.} For such a field spectrum, we find $P_{\rm dm}^{(\rm iso)}(y_0,k)=\pi^{3/2}k_*^{-3}e^{-k^2/(4k_*^2)}$. The $P_{\rm cdm}(y_0,k)$ in the above expressions is the usual CDM adiabatic power spectrum for $y_0\ll 1$, and $\mathcal{T}^2_{\rm cdm}(y,y_0,k)$ represents the growth of the power spectrum since that time.

\begin{figure*}[t!]
\includegraphics[width=\textwidth]{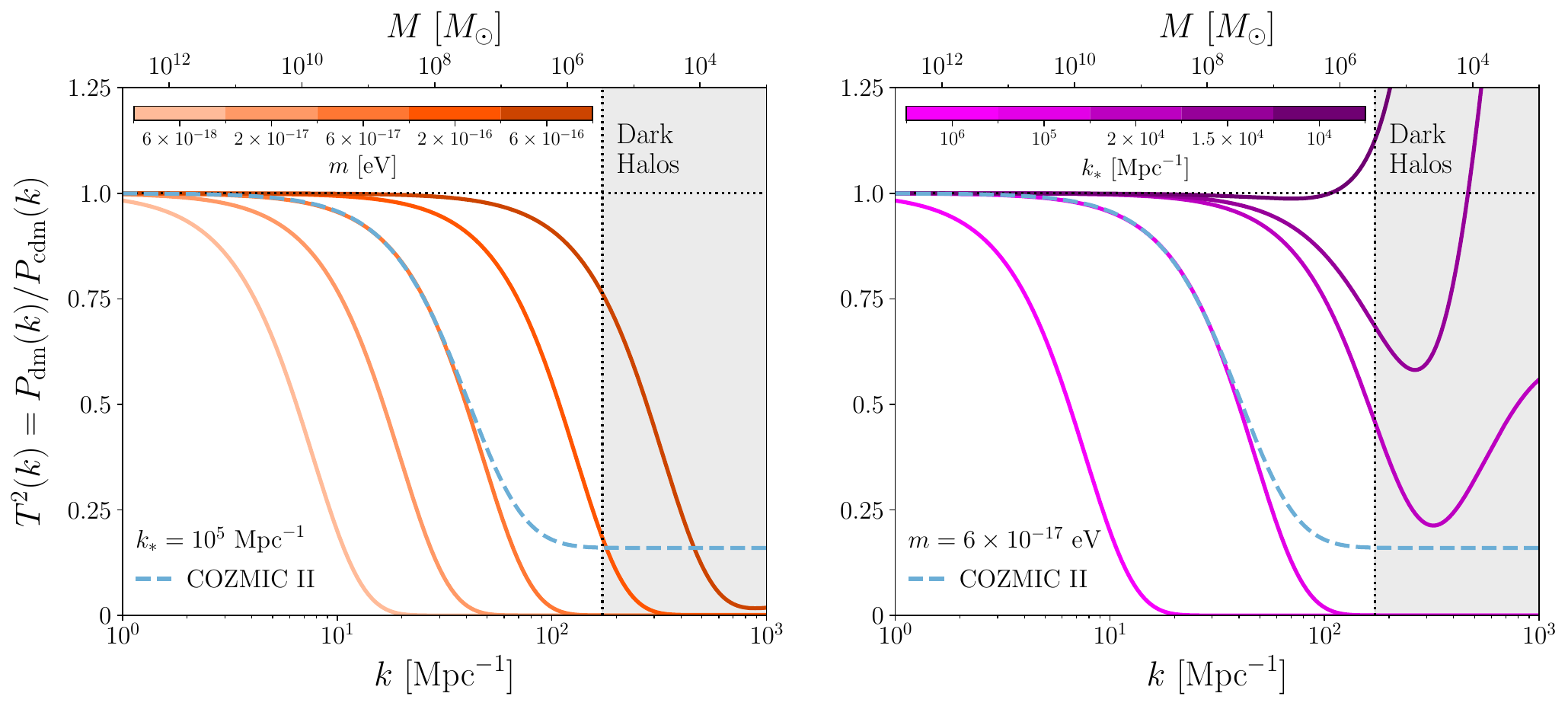}
\vspace{-6mm}
\caption{Ratio of the linear matter power spectrum for our scenario, in which DM production imprints a free-streaming cutoff and a Poisson enhancement, versus that in CDM. The left panel shows transfer functions at a fixed $k_*=10^5~\mathrm{Mpc}^{-1}$ for five values of $m$ (with $m$ increasing from the lightest to darkest shade); the right panel shows fixed $m=6\times 10^{-17}~\mathrm{eV}$ for five values of  $k_*$ (with $k_*$ decreasing from the lightest to darkest shade). The dashed blue line shows the square of the COZMIC II transfer function with a $40\%$ plateau height, which is ruled out at $95\%$ confidence by the MW satellite population~\citep{An241103431}; ultralight DM models with power spectra that are strictly more suppressed than this reference for $k<172~\mathrm{Mpc}^{-1}$ are ruled out. The dotted black vertical line marks $k=172~\mathrm{Mpc}^{-1}$; larger wavenumbers in the shaded gray region correspond to halo masses below the galaxy formation threshold in the presence of molecular hydrogen cooling and DM--baryon streaming motion~\citep{Nadler250304885}.}
\label{fig:T_DM}
\end{figure*}

The transfer function $T_{\rm dm}$ has the following characteristic scales determined by $k_*$ and $m$:
\begin{align}
\label{eq:kfskJ}
k_{\rm fs}^{\rm eq}\sim \frac{k_{\rm eq}}{\frac{k_*}{a_{\rm eq}m}}\frac{1}{\ln\left(\frac{a_{\rm eq} m}{k_*}\right)},\quad
k^{\rm eq}_{\rm J}\sim \frac{\sqrt{3}}{2}\frac{k_{\rm eq}}{\frac{k_*}{a_{\rm eq}m}}.
\end{align}
Here, $(k^{\rm eq}_{\rm fs})^{-1}$ is the comoving free-streaming length at matter--radiation equality. Adiabatic perturbations are erased below this length during radiation domination. During matter domination, density perturbations do not undergo significant growth below the Jeans length, $(k^{\rm eq}_{\rm J})^{-1}$, while the growth of larger-scale modes is proportional to $y$. It is worth noting that $\sigma_{\rm eq}\equiv k_*/a_{\rm eq}m$ is the effective velocity dispersion at matter--radiation equality.\footnote{Note that while $k_{\rm fs}$ changes very slowly for $y\gg 1$, $k_{\rm J}\propto \sqrt{y}$. Also note that $k_{\rm fs}$ and $k_{\rm J}$ are different from $k_{\rm j}\sim a\sqrt{mH}$ in \cite{Hu008506}.}

The expected shape of $T_{\rm dm}^2$ at $y\gg 1$ is as follows. For $k\ll k_{\rm fs}^{\rm eq}$, the transfer function approaches unity, whereas for $k\gg k_*$ it approaches zero exponentially.
Depending on the value of $k_*$ and $m$, $T_{\rm dm}^2$ can be suppressed due to free-streaming for $k\gtrsim k_{\rm fs}^{\mathrm{eq}}$ followed by an enhancement that scales as $(k/k_*)^3$.

Finally, we take the transfer function to be zero for $k>k_{\rm J}$. Although there is power at higher $k$, it is only transient Poisson noise not associated with gravitational clustering, so it should not contribute to halo formation. This truncation choice is conservative; for example, \cite{Amin250612131,Amin250320881} found that halo mass functions are predicted reasonably well if $T_{\mathrm{dm}}^2$ is truncated at the lower wavenumber of $k_{\rm J}/4$.

The above discussion assumed that $k_{\rm fs}^{\rm eq}<k_{\rm J}^{\rm eq}<k_*$. This ordering arises when $k_*\gg k_{\rm j}^{\rm eq}=k_{\rm eq}\sqrt{m/H_{\rm eq}}$. However, in the case where $k_*\rightarrow 0$ (i.e., when we recover the results from the globally misaligned cold FDM case), the free-streaming and Jeans scales $k_{\rm J}^{\rm eq},k_{\rm fs}^{\rm eq}\rightarrow \infty$, so they do not play any role. In this limit, the relevant Jeans scale becomes $k_{\rm j}^{\rm eq}$, which is the cold FDM Jeans scale of \cite{Hu008506}.

\subsection{Characteristic Behavior}
\label{sec:behavior}

Figure~\ref{fig:T_DM} shows $T^2_{\mathrm{dm}}(k;m,k_*)$ for illustrative values of $m$ and $k_*$ at $z=99$. We choose this redshift to match the initialization redshift of the COZMIC simulations, since we will map to the COZMIC II transfer function limit from \cite{An241103431}. The left panel demonstrates that, for fixed $k_*$, increasing $m$ shifts the $P(k)$ cutoff to smaller scales. This is reminiscent of cold FDM, but here the cutoff is due to free-streaming rather than the  ``quantum'' Jeans scale associated with the DM particle mass. In particular, $k_{\mathrm{fs}}^{-1}\propto (k_*/m)\ln(a_{\mathrm{eq}}m/k_*)$, so smaller values of $k_*/m$ yield smaller free-streaming lengths.

The right panel of Figure~\ref{fig:T_DM} shows $T^2_{\mathrm{dm}}(k;m,k_*)$ for $m=6\times 10^{-17}~\mathrm{eV}$ and five values of $k_*$. As $k_*$ decreases, the Poissonian white-noise term in Equation~(\ref{fig:T_DM}), which scales as $k_{*}^{-3}$, becomes more important. This enhancement competes with the free-streaming cutoff and can overwhelm it for sufficiently small $k_*$, resulting in $P(k)$ enhancement (rather than suppression) relative to CDM on MW satellite scales. This regime is illustrated by the darkest magenta line in the right panel of Figure~\ref{fig:T_DM}, with $k_*=10^4~\mathrm{Mpc}^{-1}$.

The gray band shows wavenumbers $k>172~\mathrm{Mpc}^{-1}$. These wavenumbers correspond to halos below the galaxy formation threshold, which we conservatively identify as a peak mass of $10^6~M_{\mathrm{\odot}}$~\citep{Nadler250304885}, where we associate wavenumbers with masses in linear theory, following~\cite{Nadler190410000,Nadler241003635}. We will only require transfer functions for ruled-out DM models to be more suppressed than the COZMIC result for $k<172~\mathrm{Mpc}^{-1}$, since excess power on smaller scales does not affect MW satellite abundances~\citep{Nadler250716889}.

\section{Limits}
\label{sec:results}

\begin{figure}[t!]
\centering
\hspace{-6mm}
\includegraphics[width=0.5\textwidth]{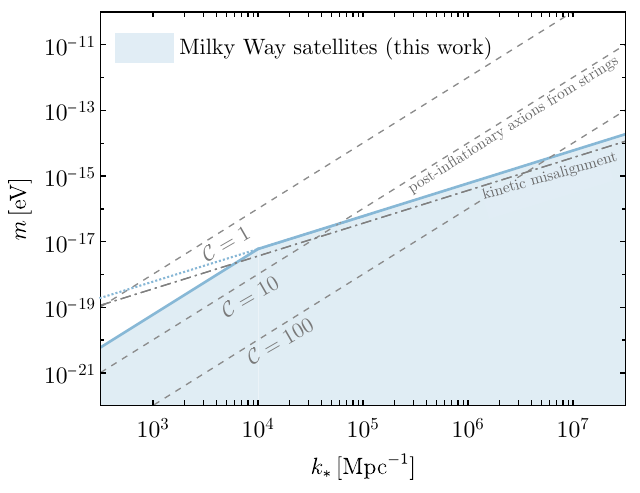}
\vspace{-5mm}
\caption{Limits on the DM particle mass, $m$, versus the comoving wavenumber of field modes that dominate the DM density, $k_*$, based on DES and PS1 MW satellite observations (solid blue line; the shaded region is excluded). The dotted line shows an extrapolation of the $m\sim k_*$ scaling to lower $k_*$ as a visual guide; for $k_*< 10^4~\mathrm{Mpc}^{-1}$, the scaling of our mass bound steepens to $m\sim k_*^2$ {(see the main text for details). Gray curves show representative theoretical relationships between $m$ and $k_*$ motivated by the production mechanisms discussed at the end of Section~\ref{sec:results}.
}} 
\label{fig:constraint}
\end{figure}

We compute lower limits on $m$ as a function of $k_*$ by requiring that $T^2_{\mathrm{dm}}(k; m,k_*)$ is strictly more suppressed than the square of the ruled-out COZMIC II transfer function with a plateau height of $40\%$ for $k<172~\mathrm{Mpc}^{-1}$.\footnote{The COZMIC II transfer function with $0.6$ plateau height is unconstrained in the \cite{An241103431} MW satellite analysis; thus, our bounds represent the strongest constraint on warm ultralight DM that can be obtained at $95\%$ confidence based on COZMIC II results. Meanwhile, adopting the constrained $0.2$ plateau height model only weakens our bounds by $\approx 25\%$.} This procedure yields the constraints shown by the solid blue line in Figure~\ref{fig:constraint}, which can be written as
\begin{equation}
\begin{cases}
    m>6\times 10^{-18}~\mathrm{eV}\times\left(k_*/10^4\,{\rm Mpc}^{-1}\right),~\ k_*> 10^4~\mathrm{Mpc}^{-1},\\
    m > 6\times 10^{-18}~\mathrm{eV}\times\left(k_*/10^4\,{\rm Mpc}^{-1}\right)^2,~\ k_*< 10^4~\mathrm{Mpc}^{-1}.
\end{cases}
\end{equation}

At this transition value of $k_*$, the Poisson enhancement begins to affect scales relevant for MW satellites. In particular, for $k_*> 10^4\,\Mpc^{-1}$, the $T^2_{\rm dm}(k)$ cutoff in our scenario precisely matches WDM on MW satellite scales, given appropriate choices of $m$. In this regime, we only probe the ratio $m/k_*$, which sets the free-streaming scale, so we obtain a limit that scales as $m\propto k_*$.

For $k_*< 10^4~\mathrm{Mpc}^{-1}$, the white-noise enhancement of the transfer function, $T_{\rm dm}^2(k)\propto (k/k_*)^3$, can exceed the ruled-out COZMIC transfer function even if we reduce $m\propto k_*$. By imposing that the largest enhancement, $\sim (k_{\rm J}/k_*)^3$, must remain below the COZMIC transfer function up to $172\,\Mpc^{-1}$, we obtain a bound on the mass $m$ that is weaker than expected from the $m\propto k_*$ scaling. In the $m$--$k_*$ plane, the bound lies along the curve $k_{\rm J}/k_*\propto m/k_*^2={\rm const}$. 
This heuristic argument explains the shape of the limits we obtain for $k_*<10^4~\mathrm{Mpc}^{-1}$. If we assume that this Jeans-related cutoff occurs at $k_{\rm J}/4$ rather than $k_{\rm J}$, based on the results of \cite{Amin250320881}, our bounds strengthen by $\approx 30\%$ for $k_*<10^3~\mathrm{Mpc}^{-1}$ and are largely unaffected at larger $k_*$.

Appendix~\ref{sec:t_saturate} illustrates transfer functions that saturate the constraint in the $k_*< 10^4~\mathrm{Mpc}^{-1}$ regime. We plot our limit down to $k_*=10^2~\mathrm{Mpc}^{-1}$ because independent data from the Ly$\alpha$ forest confidently establish $k_*>10^2~\mathrm{Mpc}^{-1}$ (see Section~\ref{sec:discussion}). We note that the COZMIC II simulations sample a finite set of transfer functions, but that this has a negligible impact on our results; for example, shifting the cutoff wavenumber of the reference COZMIC II transfer function by $10~\mathrm{Mpc}^{-1}$ (corresponding to the spacing of the COZMIC II simulation grid) in either direction only alters our mass bounds by $\approx 15\%$.

In Figure~\ref{fig:constraint}, we also demonstrate the potential impact of our bound on some representative models. The theoretically motivated gray (dashed and dashed--dotted) curves in the $m$--$k_*$ plane are provided for illustrative purposes, and not as exhaustive scans of the underlying theories. The gray dashed lines with $m\propto k_*^2$ are obtained, for example, for postinflationary axion production (see \citealt{Vaquero180909241,Buschmann190600967,Gorghetto200704990,Saikawa:2024bta}). In these theories, $k_*= \mathcal{C} m a_m$ where $H(a_m)\sim m$. Here, $\mathcal{C}=\mathcal{O}(1)$ is expected for the case where axion strings do not play a significant role, whereas \cite{Gorghetto:2025uls} quote $\mathcal{C}= \mathcal{O}(10)$ as their inferred value from simulations including strings. Another general possibility is $k_*=ma_*$, where $a_*\gg a_m$, which can be achieved by a variety of mechanisms, including those where $a_*$ is approximately independent of $m$. For example, for axions produced by kinetic misalignment~\citep{Co:2019jts} with constant barrier height and temperature-independent mass~\citep{Eroncel:2022vjg}, we obtain a representative dashed--dotted curve $m\propto k_*$, again illustrating the reach of our bound. We reiterate that, in general, $m\propto k_*^\alpha$, where $\alpha$ and the coefficient of proportionality depend on details of the scenario (and both need not be constant for all $k_*$).

\section{Discussion}
\label{sec:discussion}

\subsection{Robustness of Limits}

For $k_*> 10^4~\mathrm{Mpc}^{-1}$ the transfer functions of DM models along our constraint are very similar to ruled-out thermal-relic WDM models on all scales relevant for MW satellites. At lower $k_*$, our mass bounds weaken because the transfer function becomes less similar to WDM. In this regime, our procedure only constrains transfer functions with sufficiently small $m$, such that the white-noise enhancement does not exceed the COZMIC transfer function up to $k_{\mathrm{threshold}}=172~\mathrm{Mpc}^{-1}$. For $k_*\approx 10^2~\mathrm{Mpc}^{-1}$, the Jeans scale cuts off the transfer function before it crosses above the COZMIC limit. Adjusting this threshold to $k_{\mathrm{threshold}}=80~\mathrm{Mpc}^{-1}$ (corresponding to a peak mass scale of $10^7~M_{\mathrm{\odot}}$) only strengthens our ultralight DM mass limits by $\approx 10\%$ for $k_*\lesssim 10^3~\mathrm{Mpc}^{-1}$.

Thus, the transfer functions of all models along our constraint boundary are strictly more suppressed than the ruled-out COZMIC transfer function. The DM models we rule out therefore suppress satellite galaxy abundances strictly more than WDM models that have been well studied, yielding a lack of observable MW satellites that is inconsistent with DES and PS1 data. Note that the MW satellite limits we use incorporate uncertainties in the galaxy--halo connection and the properties of the MW system while accounting for the spatial anisotropy and limited detectability of MW satellites~\citep{Nadler191203302,Nadler200800022}. Our results are conservative and robust in this sense. However, we make the simplifying assumption that the form of the galaxy--halo connection in the warm ultralight DM models we consider is unchanged relative to CDM, and that degeneracies between galaxy--halo connection and DM model parameters are consistent with COZMIC results. These assumptions, which are, respectively, consistent with hydrodynamic simulation results in other beyond-CDM models (e.g., \citealt{Despali250112439}) and with previous non-CDM galaxy--halo connection inferences~\citep{Nadler200800022}, represent natural areas for follow-up work in our scenario.

Finally, we note that several other small-scale structure probes yield $P(k)$ limits comparable to the MW satellite population. For example, analyses of the Ly$\alpha$ forest flux power spectrum \citep{Viel1308804,Irsic179602,Irsic230904533}, strong lensing flux ratio statistics \citep{Gilman190806983,Hsueh190504182,Keeley240501620}, stellar stream perturbations \citep{Banik191102663}, and combinations thereof \citep{Enzi201013802,Nadler210107810} yield transfer function constraints near to or stronger than the COZMIC limit we have adopted. Thus, the methods developed here can be applied to other observables to obtain similar DM mass bounds.

\subsection{Comparison to Previous Results}

We now compare to previous limits on the DM particle mass in our scenario. \cite{Amin221109775} obtained $m\gtrsim 10^{-19}~\mathrm{eV}$ from the Ly$\alpha$ forest. A bound on the mass, rather than on $m/k_*$, is obtained because these authors use both (1) a lack of white-noise enhancement for $k<k_{\rm obs}\sim 10\,\Mpc^{-1}$ (which implies $k_* \gtrsim 10^3~\mathrm{Mpc}^{-1}$), and (2) an absence of a free-streaming cutoff for $k\lesssim k_{\rm obs}$. That is, $k_{\rm fs}\gtrsim k_{\rm obs}$, which implies $m/k_*\gtrsim 10^{8}$. Together with (1), this yields the stated bound on the mass. In the $m$-$k_*$  plane, (1) and (2) yield a $\Vone$-shaped allowed region, the boundary of which roughly corresponds to a vertical line at  $k_*\sim 10^3\,\Mpc^{-1}$ joined to the dotted line shown in Figure~\ref{fig:constraint}. The corner of this region (at $k_*=10^{3}\,\Mpc^{-1}$) then yields the lower bound on the mass. In the present Letter, we obtain $m>6\times 10^{-20}~\mathrm{eV}$ at $k_*=10^3\,\Mpc^{-1}$, which is slightly less stringent than the \cite{Amin221109775} bound, because our constraint weakens in this regime to avoid white-noise enhancement on MW satellite scales.

In \cite{Long241214322},  the authors used eBOSS DR14 Ly-$\alpha$ forest data, which extends to $k_{\rm obs}\sim {\rm 3}\,\Mpc^{-1}$,  and also obtained an approximately $\Vone$-shaped allowed region in the $m$-$k_*$ plane. Their limits on the mass are slightly weaker than those of \cite{Amin221109775}, who used $k_{\rm obs}\sim 10\,\Mpc^{-1}$ based on the discussion in \cite{Irsic191111150}. Their limits are also weaker by a factor of a few than the ones derived in the present Letter for $k_*> 10^4~\mathrm{Mpc}^{-1}$.

For misalignment production of cold FDM, which does not feature free-streaming, bounds on $m$ weaken compared to our limits. For example, \cite{Rogers200712705} derived $m>2\times 10^{-20}~\mathrm{eV}$ using the Ly$\alpha$ forest and \cite{Nadler241003635} derived $m>1.4\times 10^{-20}~\mathrm{eV}$ using MW satellites. Both limits are based on FDM Jeans suppression alone, corresponding to $k_*\rightarrow 0$ in our scenario. In this limit, isocurvature density fluctuations cannot be observationally constrained because $k_*$ is so low that fluctuations only occur on lengthscales and time scales larger than the present-day Hubble scale.

Meanwhile, \cite{Irsic191111150} analyzed a postinflationary DM production scenario with $k_*\sim a(t)m$ when $H(t)=m$. These authors used Ly$\alpha$ forest data but did not model the free-streaming cutoff, finding $m>2\times 10^{-17}~\mathrm{eV}$. In this scenario, the relation between $k_*$ and $m$ yields a bound on $m$ from the observed lack of white-noise enhancement. Also see related recent work by \cite{Chathirathas:2025aan}.

Lastly, our constraint can be translated to a limit on the redshift at which DM becomes nonrelativistic, $z_{\mathrm{nr}}=m/k_*>9.5\times 10^7$ for $k_*> 10^4~\mathrm{Mpc}^{-1}$. \cite{Das2010101137} placed constraints on the redshift $z_T$ when DM is instantaneously produced from either a free-streaming or self-interacting relativistic (neutrino-like) species. Their bounds, which are also derived by mapping to the WDM constraint from \cite{Nadler200800022}, are weaker than our limit on $\znr$ by a factor of $\approx 20$. However, we caution against a detailed comparison because $\znr$ and $z_T$ are set by different underlying physics.

\subsection{Future Prospects}

Future data may improve our bounds in multiple ways. First, observing more satellite galaxies (even at a fixed halo mass) would strengthen constraints on the lack of free-streaming at fixed $k$ if the data are consistent with CDM. For example, \cite{Nadler240110318} showed that complete observations of two satellite populations (e.g., the MW and M31) yields $m_{\mathrm{WDM}}\gtrsim 20~\mathrm{keV}$, assuming underlying CDM subhalo abundances. In our scenario, this forecast translates to a limit of $m>3\times 10^{-16}~\mathrm{eV}$ at $k_*=10^5~\mathrm{Mpc}^{-1}$, improving our fiducial bound by a factor of $\approx 5$ at this value of $k_*$.

On the other hand, if small-scale $P(k)$ suppression and enhancement can both be probed, then $m$ can be constrained independent of $k_*$, similar to the Ly$\alpha$ bound from \cite{Amin221109775}. In this regime, lower limits on $m$ scale quadratically with the wavenumber of the smallest scale probed, which is set by the minimum halo mass that contributes to a given observable. Halos with peak masses below the galaxy formation threshold are therefore a particularly valuable probe of the DM particle mass in our scenario, since they probe $P(k)$ on scales $k\gtrsim 200~\mathrm{Mpc}^{-1}$. These systems can be detected using strong gravitational lensing \citep[e.g.,][]{Nierenberg230910101,Powell251007382} or stellar stream perturbations \citep[e.g.,][]{Banik191102662,Nibauer251002247}.

Recently, \cite{Amin:2025nxm} studied a case where warm ultralight DM is a fraction of the total DM. In this case, the severity of $P(k)$ suppression due to free-streaming scales with the fraction of the warm component. Furthermore, the shot noise enhancement in the subdominant warm component can seed structure in the dominant cold component. For sufficiently small fractions and essentially ignoring free-streaming, \cite{Gorghetto:2025uls} obtained constraints on a subdominant postinflationary axion using small-scale structure data. We leave the use of MW satellites to constrain subdominant fractions of warm ultralight DM to future work.

\section{Conclusions}
\label{sec:conclusion}

This study generalizes lower limits on the DM particle mass $m$ using MW satellite galaxy abundances by casting them as a function of $k_*$, the comoving length scale on which there are initially $\mathcal{O}(1)$ fluctuations in the DM field; thus, $k_*$ is a typical scale of the comoving DM field gradient, and hence its characteristic comoving momentum.  Our analysis conservatively yields $m>6\times 10^{-18}~\mathrm{eV}\times (k_*/10^4\,{\rm Mpc}^{-1})$ for $k_*> 10^4~\mathrm{Mpc}^{-1}$, at $95\%$ confidence; these limits weaken to $m>6\times 10^{-18}~\mathrm{eV}\times (k_*/10^4\,{\rm Mpc}^{-1})^2$ for $k_*< 10^4~\mathrm{Mpc}^{-1}$ to avoid linear matter power spectrum enhancement on scales that affect MW satellite galaxy abundances. 

These limits hold if the entire DM density is produced by any process with a finite correlation length $\sim k_*^{-1}$. For example, if the DM is produced at a specific epoch, this correlation length can be set by the horizon size at that time due to causality. More generally, our limits apply to many postinflationary DM production mechanisms, as well as some scenarios in which DM is produced during inflation. In all of these scenarios, we assume that there is no significant homogeneous mode in the DM field.

We have argued that our bounds will rapidly improve as the statistical precision of small-scale structure data increases, and as future astrophysical observations probe even lower-mass halos. Our results imply that constraining both enhancement and suppression of small-scale power relative to CDM will be critical to break the degeneracy between the DM mass and its characteristic momentum. These same measurements will also test whether DM is indeed produced with a peaked field power spectrum, yielding new insights into fundamental physics.

\section*{Acknowledgments}

We are grateful to the referee for constructive comments, and to Mudit Jain, Andrew Long, Mehrdad Mirbabayi, Moira Venegas, and Huangyu Xiao for helpful discussions. We thank the Texas APS meeting for bringing M.A.A.\ and R.H.W.\ together in person, which initiated discussion leading to this work.

This material is based upon work supported by the National Science Foundation (NSF) under grant No.\ 2509561 (E.O.N. and A.B.) and No.\ 2407380 (V.G.). V.G.\ also acknowledges support from NSF CAREER grant No.\ PHY-2239205, the Research Corporation for Science Advancement under the Cottrell Scholar Program, the IBM Einstein Fellowship at the Institute for Advanced Study, and grant 63667 from the John Templeton Foundation. Any opinions, findings, and conclusions or recommendations expressed in this material are those of the author(s) and do not necessarily reflect the views of the NSF, nor of the the John Templeton Foundation. M.A.A.\ is supported by a US\ Department of Energy (DOE) grant DE-SC0010103. This work received support from the US\ DOE under contract No.\ DE-AC02-76SF00515 to SLAC National Accelerator Laboratory and from the Kavli Institute for Particle Astrophysics and Cosmology.

This work was performed in part at the Aspen Center for Physics, which is supported by National Science Foundation grant PHY-2210452. This research was supported in part by grant NSF PHY-2309135 and the Gordon and Betty Moore Foundation grant No.\ 2919.02 to the Kavli Institute for Theoretical Physics (KITP).

\bibliographystyle{yahapj2}
\bibliography{references_new}

\appendix

\section{Time Evolution of the Power Spectrum}
\label{App:PS}
The detailed derivation of the expressions below can be found in \cite{Amin250612131}. We state the results below as relevant for our calculations in the main body of the text in a self-contained manner.

The time evolution of the power spectrum of the DM density contrast is
\Beq
\label{eq:MainResultPS}
    P_{{\rm dm}}(y,k) &= \underbrace{{P}^{(\mathrm{ad})}_{{\rm dm}}(y_0,k) \left[\mathcal{T}_k^{(\mathrm{ad})}(y,y_0)\right]^2 \vphantom{\int_{y_0}^y}}_{\text{adiabatic IC + evolution}} +
    \underbrace{P^{(\mathrm{iso})}_{\rm dm}(y_0,k) \left[\mathcal{T}_k^{(\rm iso)}(y,y_0)\right]^2}_{\text{isocurvature IC + evolution}},
\Eeq
where the adiabatic and isocurvature evolution functions\footnote{In our context, the isocurvature Poisson contribution is generated after inflation and is not correlated with the adiabatic initial conditions from inflation.} are given by
\Beq \label{eq:MainResultT}
\mathcal{T}_k^{({\rm iso})}(y,y_0) &=
 \ml[1 + 3 \int_{y_0}^y \! \frac{\dl y'}{\sqrt{1+y'}} \mathcal{T}^{(\mathrm{b})}_k(y, y')\mathcal{T}^{(\mathrm{c})}_k(y,y')\mr]^{1/2}\,,\\
\mathcal{T}^{(\mathrm{ad})}_k(y,y_0) &= \mathcal{T}^{(\mathrm{a})}_k(y,y_0) + \frac{1}{2} \frac{\dl\ln(P_{{\rm cdm}}(y_0,k))}{\dl\ln(y_0)} \sqrt{1+y_0} \, \mathcal{T}^{(\mathrm{b})}_k(y,y_0)\,.
\Eeq
Here, $y_0\ll 1$ is at an initial ``time" when all wavenumber-$k$ modes of interest are subhorizon, and the field modes of interest are nonrelativistic; the initial conditions (IC) are specified at that time. We discuss this further at the end of this section. The adiabatic IC is $P_{\rm cdm}(y_0,k)\approx 36P_{\mathcal{R}}(k)\left[3+\ln\ml(0.15k/k_{\mathrm{eq}}\mr) - \ln\ml(4/y_0\mr)\right]^2$, with $k^3/(2\pi^2)P_\mathcal{R}(k)\approx 2\times 10^{-9}$.

The three different 
$\mathcal{T}_k^{(\mathrm{a}, \mathrm{b}, \mathrm{c})}$ in the above expressions are determined by the following Volterra equations:\footnote{Heuristically, $\mathcal{T}_k^{(\mathrm{a,b})}$ describe the evolution of bulk density and velocity perturbations, respectively, while $\mathcal{T}_k^{(\mathrm{c})}$ is related to the evolution of Poisson fluctuations.}
\Beq
\label{eq:Ty}
    \mathcal{T}^{(i)}_k(y,y') &= {\mathcal{T}}^{\mathrm{fs}\,(i)}_k(y,y')+\frac{3}{2}\int_{y'}^y \frac{\dl y''}{\sqrt{1+y''}}{\mathcal{T}}^{\mathrm{fs}\,(\mathrm{b})}_k(y,y'') {\mathcal{T}}^{(i)}_k(y'',y')\,,\quad i=\mathrm{a,b,c}.
\Eeq
Solving these Volterra equations requires a specification of the the free-streaming kernels, ${\mathcal{T}}^{\mathrm{fs}\,(\mathrm{a,b,c})}_k$, which can be calculated based on initial field power spectra, which we take to have a Maxwell--Boltzmann shape:
\Beq
f_0({\bm q})=\frac{(2\pi)^{3/2}}{k_*^{3}}e^{-q^2/2k_*^2}.
\Eeq
Here, all momenta/wavenumbers are comoving. Note that $k_*\rightarrow 0$ leads to $f_0({\bm q})\rightarrow (2\pi)^3\delta_{\rm D}({\bm q})$, a Dirac delta function. This is the limit in which we recover the globally misaligned cold FDM case. 

The free-streaming kernels are given by
\Beq
    \mathcal{T}^{\mathrm{fs}\,(\mathrm{a})}_k(y,y') &= \cos[\gamma\alpha_k^2\mathcal{F}(y,y')]\int_{\bm q} f_0(q) \exp\ml[-i\hat{{\bm q}} \cdot \hat{{\bm k}}\frac{q}{k_*}\alpha_k\mathcal{F}(y,y')\mr]=\cos[\gamma\alpha_k^2\mathcal{F}(y,y')]\exp\ml[-\alpha_k^2\mathcal{F}^2(y,y')/2\mr],
    \\
    \mathcal{T}^{\mathrm{fs}\,(\mathrm{b})}_k(y,y') &= \frac{1}{\gamma \alpha_k^2} \sin[\gamma \alpha_k^2\mathcal{F}(y,y')] \int_{\bm q} f_0(q) \exp\ml[-i\hat{{\bm q}}\cdot \hat{{\bm k}}\frac{q}{k_*}\alpha_k\mathcal{F}(y,y')\mr]=\frac{1}{\gamma\alpha_k^2}\sin[\gamma\alpha_k^2\mathcal{F}(y,y')]\exp\ml[-\alpha_k^2\mathcal{F}^2(y,y')/2\mr],
    \\
    \mathcal{T}^{\mathrm{fs}\,(\mathrm{c})}_k(y,y') &= \frac{\int_{\bm q}  f_0(|{\bm q}+{\bm k}/2|)f_0(|{\bm q}-{\bm k}/2|) \exp\ml[-i\hat{{\bm q}}\cdot \hat{{\bm k}}\frac{q}{k_*}\alpha_k \mathcal{F}(y,y')\mr]}{P^{(\rm iso)}_{\rm dm}(y_0,k)}=\exp\ml[-\alpha_k^2\mathcal{F}^2(y,y')/4\mr].
\Eeq
Here, $\mathcal{F}(y,y') = \ln\ml[(y/y')(1+\sqrt{1+y'})^2 / (1+\sqrt{1+y})^2\mr]$ captures the functional dependence of the comoving distance traveled by a field fluctuation during the time interval between $y'$ and $y$, and we define the parameters
\Beq
    \alpha_k \equiv \sqrt{2}\frac{k}{k_{\mathrm{eq}}}\frac{k_*}{a_{\mathrm{eq}} m},
    \qquad
    \gamma \equiv \frac{1}{2\sqrt{2}}\frac{a_{\mathrm{eq}} m}{k_*}\frac{k_{\mathrm{eq}}}{k_*}.
\Eeq

There are two relevant scales in the evolution of the density power spectrum. The first is the free-streaming scale
\Beq
    \label{eq:kfs}
    k_{\mathrm{fs}}(y) \equiv \left[\frac{1}{a_{\rm eq}}\int_{y_0}^y \dl y' \sigma(y') \frac{dy'}{{y'}^2 H(y')}\right]^{-1} ,
    \qquad \text{with} \quad
    \sigma(y) \equiv \frac{1}{y}\frac{1}{a_{\rm eq}m}\left(\frac{1}{3}\int_{\bm p} p^2f_0(p)\right)^{1/2}
    = \frac{1}{y}\frac{k_*}{a_{\rm eq}m} \equiv \frac{\sigma_{\rm eq}}{y}.
\Eeq
The free-streaming length is the typical comoving distance traveled by waves from $y_0$ to $y$. It is dominated by the motion during the radiation era. For $y\gg 1$, it saturates to $k^{\rm eq}_{\mathrm{fs}}
\sim 2^{-1/2}({k_{\rm eq}}/{\sigma_{\rm eq}) \ln(1/\sigma_{\rm eq})}$
if $y_0\sim \sigma_{\rm eq}\ln (1/\sigma_{\rm eq})$ as discussed below.
Second, we define the Jeans scale
\Beq
k_{\rm J}(y)\equiv a_{\rm eq}y\frac{\sqrt{4\pi G\bar{\rho}_{\rm dm}(y)}}{\sigma(y)}\sim \frac{\sqrt{3y}}{2}\frac{k_{\rm eq}}{\sigma_{\rm eq}}.
\Eeq
This is the $k$ above which fluctuations do not cluster in the matter-dominated era. 

In terms of these scales, note that $\alpha_k\sim k/k_{\rm J}^{\rm eq}$ and $\gamma\sim k^{\rm eq}_{\rm J}/k_*$. It is worth noting that the combination $\gamma \alpha_k^2$ appearing in the trigonometric functions scales as $(k/k^{\rm eq}_{\rm j})^2$, where $k_{\rm j}^{\rm eq}\equiv \sqrt{k_* k^{\rm eq}_{\rm J}}=a_{\rm eq}\sqrt{m H_{\rm eq}}$. This is the usual FDM scale evaluated at matter--radiation equality. Note that this scale is independent of $k_*$ and survives in the $k_*\rightarrow 0$ limit (whereas $k_{\rm fs},k_{\rm J}\rightarrow \infty$).

To ensure that free-streaming starts only after the modes of interest are sufficiently subhorizon, we define the initial time $y_0$ in the following way. We want to ensure that $k_{\rm fs}^{\rm eq}$ is subhorizon when we begin our integration. That is, we require
\Beq
k^{\rm eq}_{\mathrm{fs}}\gtrsim a_0 H(a_0)\sim y_0^{-1} k_{\rm eq}/{\sqrt{2}}\Longrightarrow y_0\gtrsim \sigma_{\rm eq}\ln (1/\sigma_{\rm eq}).
\Eeq
The above estimate is heuristic; to be conservative, one should use $y_0\gtrsim\mathcal{O}[10]\times \sigma_{\rm eq}\ln(1/\sigma_{\rm eq})$. Here, we take 
\Beq
a_0 = a_{\rm{eq}}y_0 = 30\frac{k_*}{m}\ln\left(\frac{a_{\rm{eq}}m}{k_*}\right),\label{eq:matching}
\Eeq
which ensures that our constraint on $\sigma_{\rm{eq}}$ is equal to that for the $6.5~\mathrm{keV}$ WDM model ruled out by the MW satellite population~\citep{Nadler200800022}. Note that we also require $a_0>k_*/m$, which is automatically satisfied by Equation~\ref{eq:matching}.

\section{Transfer Functions along the Constraint}
\label{sec:t_saturate}

\begin{figure}[t!]
\centering
\hspace{-6mm}
\includegraphics[width=0.5\textwidth]{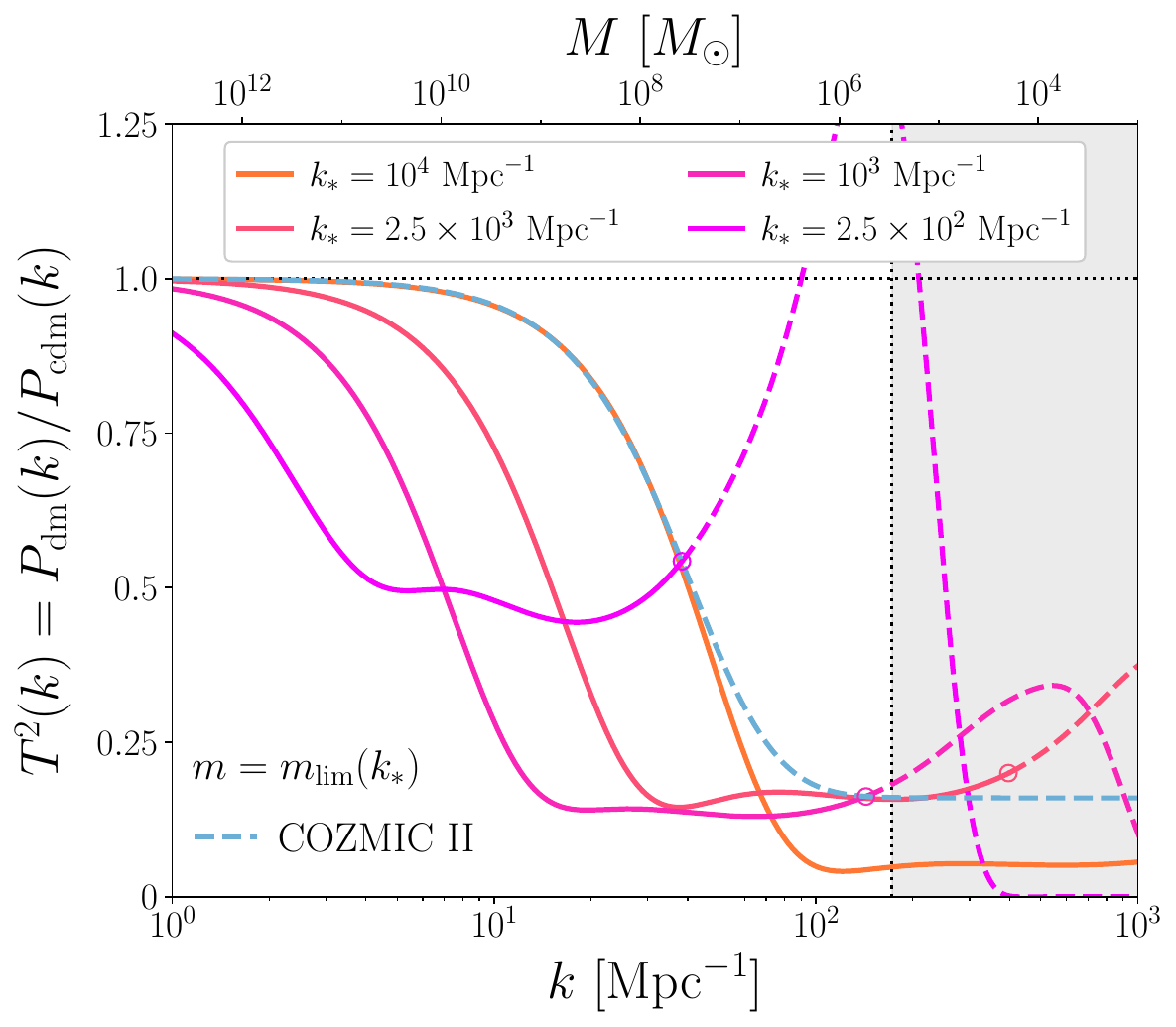}
\vspace{-2mm}
\caption{Transfer functions, evaluated along our mass constraint, for $k_*=10^4~\mathrm{Mpc}^{-1}$ (orange), $2.5\times 10^3~\mathrm{Mpc}^{-1}$ (coral), $10^3~\mathrm{Mpc}^{-1}$ (pink), and $2.5\times 10^2~\mathrm{Mpc}^{-1}$ (magenta), from right to left. Open circles show the Jeans scale $k_{\rm{J}}$; our analysis assumes that power is truncated at $k>k_{\rm{J}}$.}
\label{fig:t_dm_kJ}
\end{figure}

Figure~\ref{fig:t_dm_kJ} shows transfer functions evaluated along our constraint from Figure~\ref{fig:constraint} for $k_*\leq 10^4~\mathrm{Mpc}^{-1}$. As $k_*$ decreases, the Poisson enhancement becomes more significant on MW satellite scales. Thus, the mass bound weakens so that the Jeans scale (shown for each model by an open circle in Figure~\ref{fig:t_dm_kJ}) truncates $P(k)$ before it rises above the COZMIC II limit, which ensures that power is strictly more suppressed than the ruled-out transfer function on all scales relevant for MW satellites.

\end{document}